\documentclass[article,preprint,floatfix]{revtex4-1}
\usepackage{textcomp}
\usepackage{graphicx}
\usepackage{amssymb}
\usepackage{epsfig}
\usepackage{ifsym}
\usepackage{wasysym}
\usepackage{latexsym}

\begin{document}

\title{Direct observation of a gate tunable band-gap in electrical transport in ABC-trilayer graphene}

\author{Tymofiy Khodkov$^\dag$}
\author{Ivan Khrapach$^\dag$}
\author{Monica Felicia Craciun$^\dag$}
\author{Saverio Russo$^\dag$}
\email{Correspondence to S.Russo@exeter.ac.uk}
\affiliation{$^\dag$ Centre for Graphene Science, College of Engineering, Mathematics and Physical Sciences, University of Exeter, Exeter EX4 4QF, United Kingdom}

\begin{abstract}

Few layer graphene systems such as Bernal stacked bilayer and rhombohedral (ABC-) stacked trilayer offer the unique possibility to open an electric field tunable energy gap. To date, this energy gap has been experimentally confirmed in optical spectroscopy. Here we report the first direct observation of the electric field tunable energy gap in electronic transport experiments on doubly gated suspended ABC-trilayer graphene. From a systematic study of the non-linearities in current \textit{versus} voltage characteristics and the temperature dependence of the conductivity we demonstrate that thermally activated transport over the energy-gap dominates the electrical response of these transistors. The estimated values for energy gap from the temperature dependence and from the current voltage characteristics follow the theoretically expected electric field dependence with critical exponent $3/2$. These experiments indicate that high quality few-layer graphene are suitable candidates for exploring novel tunable THz light sources and detectors.  

\end{abstract}

\maketitle

\begin{figure}{}
\center
		\includegraphics[width=0.8\textwidth]{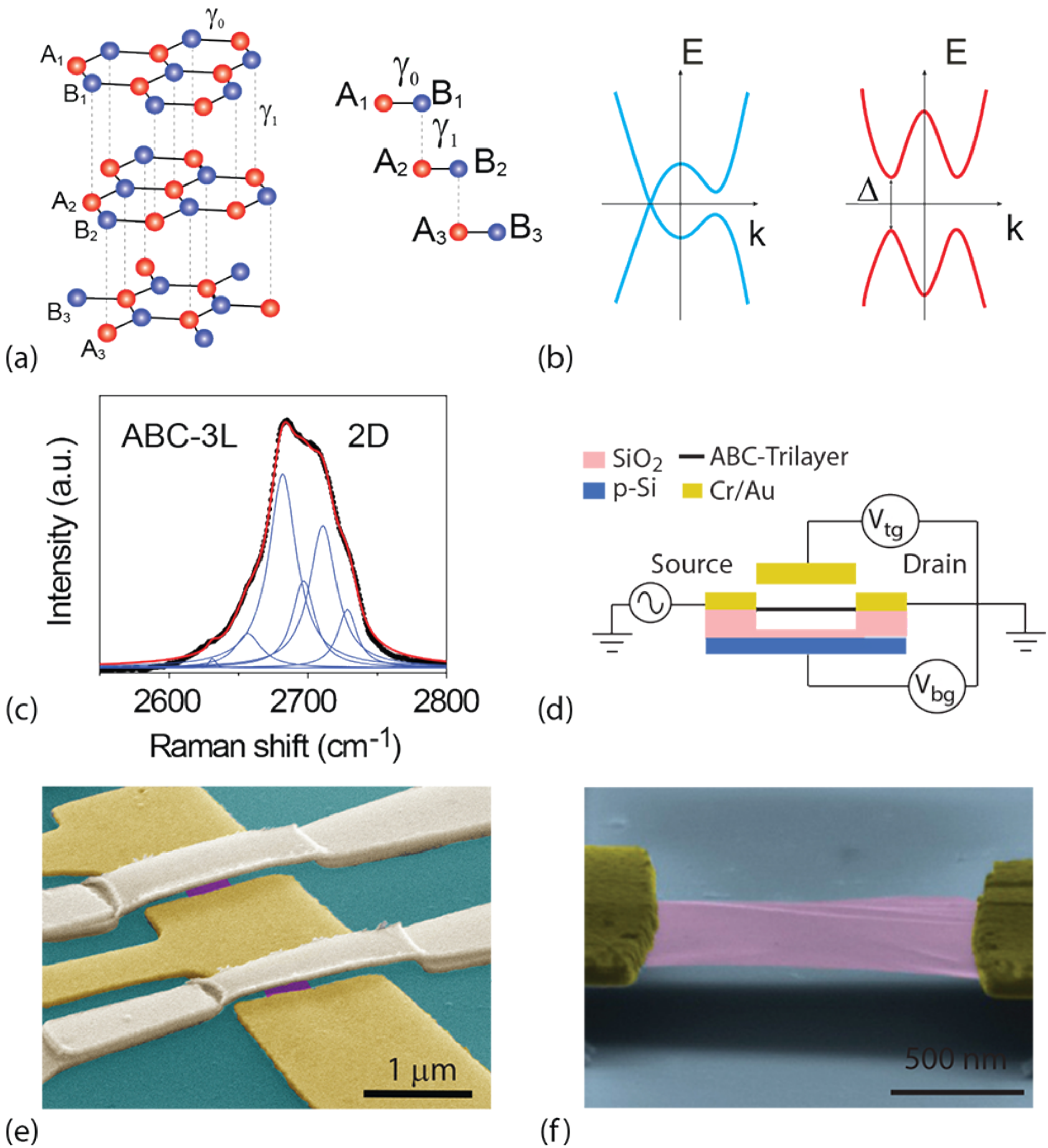}
\caption{\label{fig1} (a) Schematic crystal structure of rhombohedral (ABC-) stacked trilayer graphene (left) with the first neighbour in-plane $\gamma_0$ and out of plane $\gamma_1$ tight binding parameters highlighted. (b) Energy dispersion for ABC-trilayer with zero (left) and finite (right) external perpendicular electric field. (c) 2D-Raman peak for ABC-trilayer measured with a 532nmn laser, 5mW power and a spot size of 1.5 $\mu m$. The dots are the experimental data points, whereas the red continuous line is a fit to 6 Lorentzians (continuous blue lines). (d) Schematic cross section of a suspended double gated device and the electrical measurement configuration. Panels (e) and (f) show false color scanning electron micrographs of suspended double gated graphene devices, and a test suspended back-gated graphene structure fabricated on the same chip.}
\end{figure}

Few layer graphene (FLG) -stacked sheets of carbon atoms on a honeycomb lattice- are materials with unprecedented properties such as an electric field tunable band structure which depends on the stacking order of the layers \cite{Guinea2006,Latil2006,Aoki2007,Koshino2009,Koshino2010,
Avetisyan2010,Zhang2010,Wu2011,Kumar2011,Lui2011,Jhang2011,Bao2011,Lee2014,Craciun2009,Koshino79_2009}. Bernal (AB-) stacked bilayers and rhombohedral (ABC-) stacked trilayers open a band-gap when the energetical symmetry between the outer layers of the FLGs is broken \cite{McCann2006,Guinea2006,Latil2006,Aoki2007,Castro2007,Min2007,Zhang2009,Mak2009,Kuzmenko2009,Koshino2009,Koshino2010,Avetisyan2010,
Zhang2010,Wu2011,Kumar2011,Lui2011,Jhang2011,Oostinga2008,Taychatanapat2010,Yan2010,Xia2010,Craciun2011,Zou2014,Shioya2012}, e.g. by an external electric field (see Fig. \ref{fig1}a and b). Furthermore, ABC-stacked trilayers are also an ideal system to study competing physical aspects of the single-particle and many-body physics, with the observation of rich phase diagrams \cite{Bao2011,Lee2014}.

The unique possibility to continuously tune the band structure in FLGs simply by means of gate voltages is a novel functionality which is attracting a growing interest in fundamental science and in future graphene-based applications \cite{Craciun2011}. In particular, a gate-tunable band-gap, characteristic of few layer graphene materials, is at the core of optoelectronics. So far, only infrared spectroscopy experiments have directly measured the value of this electric field induced band-gap in both AB-bilayers \cite{Zhang2009,Mak2009,Kuzmenko2009} and ABC-trilayers \cite{Lui2011}. These experimental findings are contrasted by the lack of the direct observation of a tunable band-gap in charge transport measurements where a transport gap at a very different energy scale has been reported \cite{Oostinga2008,Taychatanapat2010,Yan2010,Xia2010}, in spite of the observation of an electric field induced insulating state indicating the opening of an energy gap \cite{Zou2014}. Proper consideration to the role of disorder induced sub-gap states can shed light onto this dichotomy between optical spectroscopy and electrical transport reports. In particular, the infrared aborption is largely dominated by the band-to-band transitions rather than transitions from the small density of disorder induced states. On the other hand, in electrical transport the hopping of electrons between sub-gap impurity states dominates the current flow hindering the observation of the intrinsic energy gap. Indeed, the characteristic energy scale (activation energy) associated to the hopping conduction mechanism is related to the ionization energy of the sub-gap states \cite{Shloski} and brings no knowledge of the intrinsic band-gap opened in the material. The direct observation of the FLG gate tunable band-gap in electrical transport experiments is still an open quest which is fuelling theoretical discussions \cite{JianLi2011}.

Here we report the direct observation of an electric field induced tunable band-gap in ABC-stacked trilayer graphene with electrical transport measurements. A systematic study of the temperature dependence of the minimum conductivity in these devices reveals that conduction is due to thermally excited charge carriers over the energy gap and not to hopping between sub-gap impurity states. We find that the magnitude of the measured band-gap depends on the intensity of the external perpendicular electric displacement with values of 4.6 meV for an average electric displacement of -110 mV/nm. The values of the energy gap estimated from the temperature dependence of the minimum conductivity correlate well to the energy range of non-linearity measured in the current-voltage (I-V) characteristics. These observations stem from the high quality of our devices which results in the absence of any hopping mechanism assisting the electrical conduction.

ABC-stacked trilayer graphene flakes are obtained by micromechanical exfoliation of pristine natural graphite onto SiO$_2$ (295 nm)/p-Si. The number of layers and the stacking order are identified by Raman spectroscopy \cite{Koh2011,LuiHeinz2011,Jhang2011}. In particular, the 2D-peak of ABC-stacked trilayers is asymmetric as shown by the relative heights of the 6 Lorentzians fitted to it (see Fig. \ref{fig1}c). Source and drain contacts are patterned by standard electron-beam lithography, metal deposition of Cr/Au (10nm/100nm) and lift-off process. The independent control of the external perpendicular electric field acting onto the trilayer, which determines the opening of a band-gap (see Fig. \ref{fig1}b), and of the charge density is readily obtained in double gated devices by changing independently the two gate voltages \cite{Craciun2011}. To achieve high quality devices, suspended double gated structures are fabricated \cite{Khodkov2012} (see Fig. \ref{fig1}d-f and supplementary information). The width and length of the top-gate electrode are much larger than the graphene channel length and width, therefore ensuring that a uniform electric field acts on the conductive channel. Typically, the charge carrier mobility in these devices is  $\mu > 25000 cm^2/Vs$  at 4.2K and at a charge concentration of $\sim 5*10^{11} cm^{-2}$, with no noticeable residual doping.

Electrical measurements were performed in a He$^3$ refrigerator in the temperature range between 300 mK and 100 K. The conductance of the suspended double gated devices was measured using a lock-in amplifier in a voltage-biased two-terminal configuration. The choice of the excitation voltage is pivotal to understand if an insulating state with an energy gap smaller than the recently reported values due to many-body states \cite{Bao2011,Lee2014} is present in ABC-trilayer. Hence we have progressively lowered the values of excitation voltage down to the lowest possible values which did not affect the reading of the maximum of resistance at any of the measured temperatures. In this way, we have also prevented heating of the electrons and the occurrence of non-equilibrium effects which could easily hinder the observation of small values of the intrinsic energy gap. 

\begin{figure}{}
\center
		\includegraphics[width=0.8\textwidth]{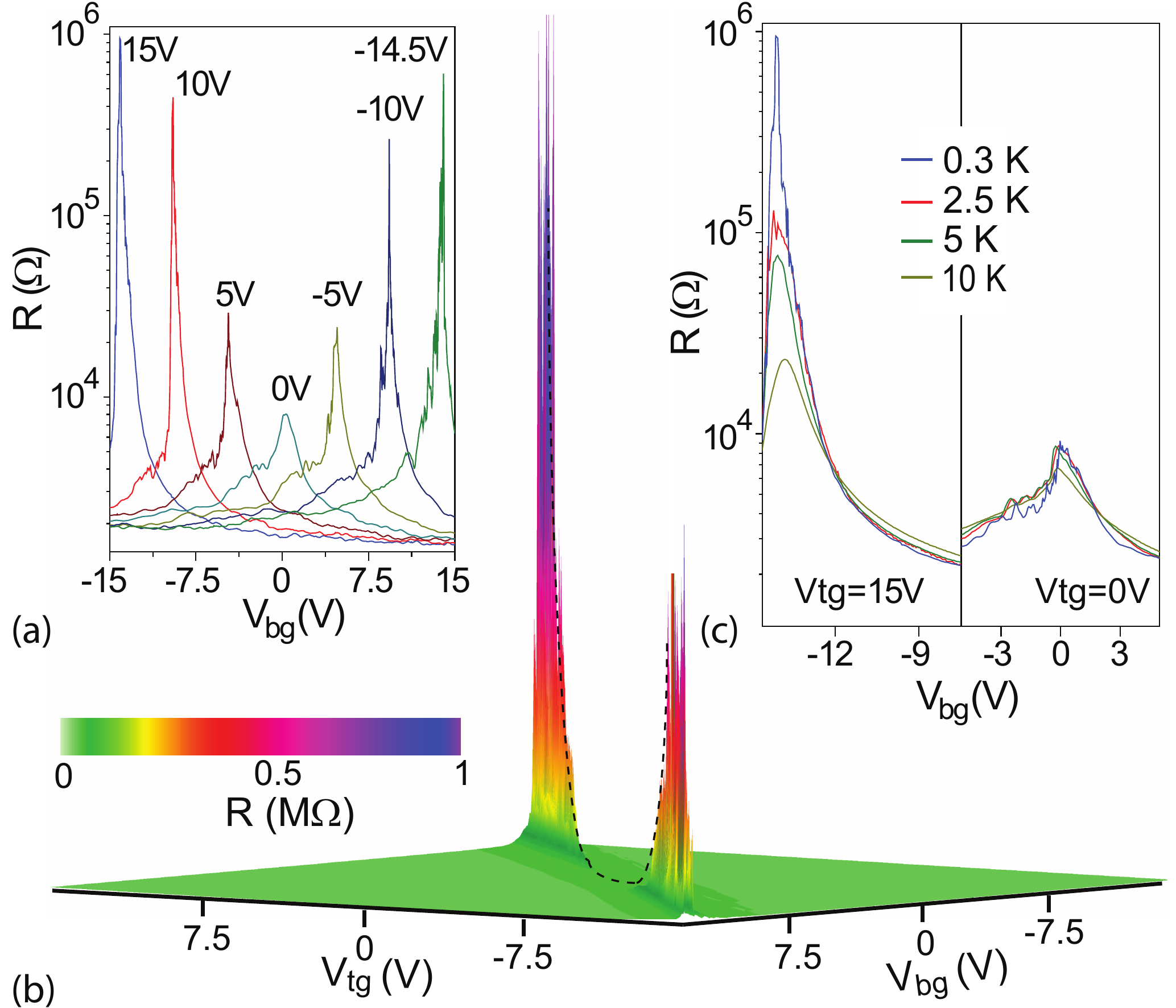}
\caption{\label{fig2} (a) Zero-bias resistance \textit{vs.} $V_{bg}$ measured at T = 0.3 K and for different fixed $V_{tg}$ as indicated in the graph. (b) 3D Plot of R \textit{vs.} $V_{bg}$ and $V_{tg}$. The dashed line is a guide line for the eyes connecting the maximum of resistance $R_{max}$. (c) $R$ \textit{vs.} $V_{bg}$ for $V_{tg}=15V$ (left) and $V_{tg}=0V$ (right) and for different temperatures.}
\end{figure}

\begin{figure}{}
\center
		\includegraphics[width=0.8\textwidth]{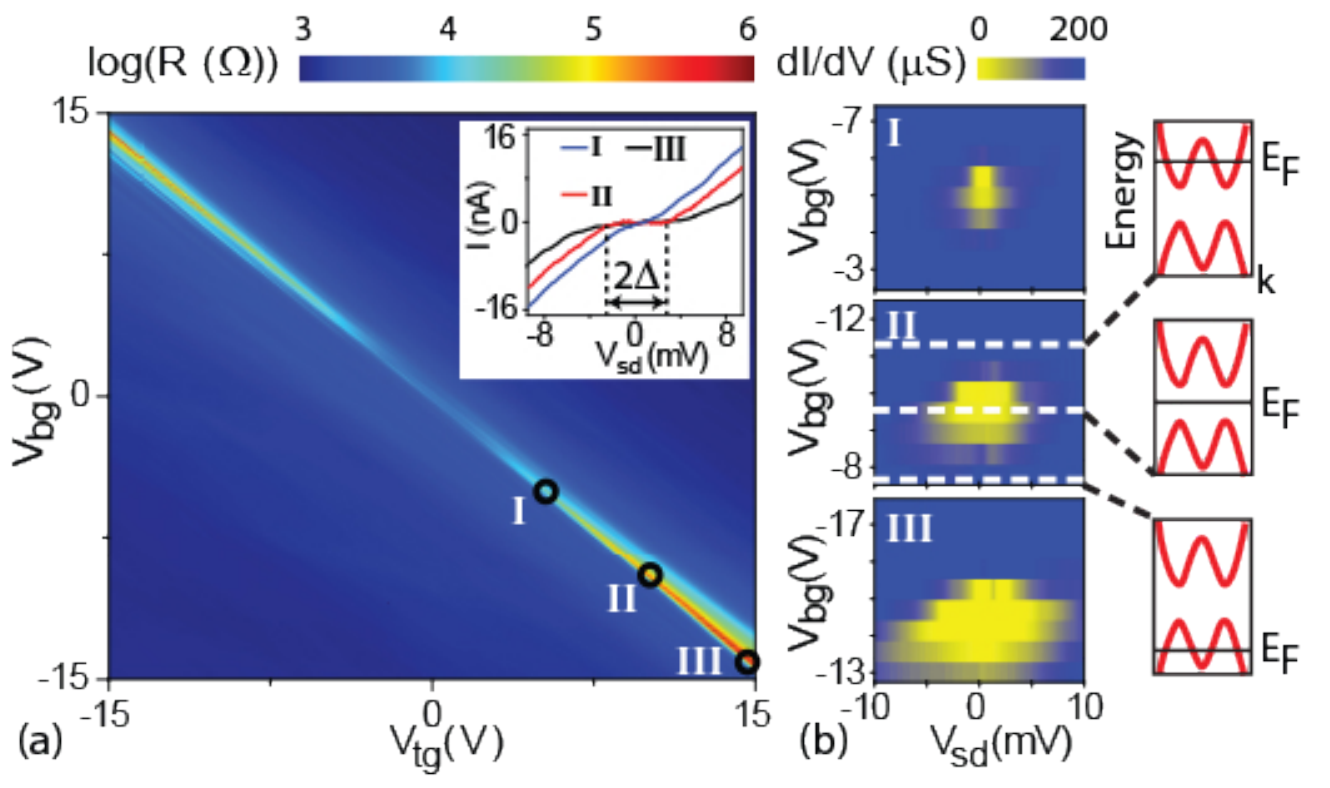}
\caption{\label{fig3} (a) Color plot of $\log (R)$ as a function of $V_{tg}$ and $V_{bg}$. The inset shows I-V characteristics at the $R_{max}$ for 3 different gate configurations highlighted in the main graph. I, II and III correspond to 0.05, 0.08 and 0.13 V/nm electric displacement. The dashed lines in the inset highlight the non-linear energy range assigned to the intrinsic gap ($\Delta$=3mV for D=0.08 V/nm). The differential conductance $dI/dV$ \textit{versus} $V_{sd}$ and $V_{bg}$ for fixed $V_{tg}$ corresponding to I, II and III is plotted in panel (b). The sketches on the right hand side show a schematic low-energy band structure of ABC-trilayer and the position of the Fermi level ($E_{F}$) for the three different values of $V_{bg}$ indicated by the white dashed lines.}
\end{figure}

Figure \ref{fig2}a and b show the resistance of a suspended double gated ABC-trilayer (channel width and length of 500nm) measured at 300 mK as a function of back-gate voltage ($V_{bg}$) for different fixed top-gate voltages ($V_{tg}$). We define the average external perpendicular electric displacement as $D = (D_{bg} + D_{tg})/2$ with $D_{bg} d_{a} + D_{bg} d_{SiO_2}/\varepsilon= V_{bg}$ and $D_{tg} d_{tg} = V_{tg}$ ($d_{tg} = 150nm$, $d_{a} = 85nm$, $d_{SiO_2}=215nm$ and $\varepsilon = 3.9$ for SiO$_2$). Upon increasing $D$, the maximum of resistance ($R_{max}$) increases from 7.9 KOhm up to 1 MOhm and shifts its position in the gate-voltages, see Fig. \ref{fig2}b. This insulating state is at least ten times more resistive than the recently reported insulating state due to many-body effects \cite{Bao2011,Lee2014} and its dependence on the voltage applied to the two gates is summarized in the color plot of Fig. \ref{fig3}a. We find that $R_{max}$ moves linearly as a function of $V_{bg}$ and $V_{tg}$, as expected for constant capacitive coupling to the two gates. At the same time, the temperature dependence of $R$ \textit{versus} $V_{bg}$ shows that the larger is $D$ the more pronounced is the temperature dependence of $R_{max}$, see Fig. \ref{fig2}c. Both the increase of $R_{max}$ as a function of $D$ and its pronounced temperature dependence at large $D$ suggest the opening of an electric-field induced band-gap due to the breaking of the inversion symmetry between the outer layers of ABC-trilayer graphene \cite{Koshino2009,Koshino2010,Avetisyan2010,Zhang2010,Wu2011,Kumar2011,Lui2011}.

Another indication of an electric field induced band-gap comes from the current voltage characteristics. The inset in Fig. \ref{fig3}a shows I-V curves for three different perpendicular electric fields configurations highlighted in the color plot. In all cases, the I-V curves display an insulating state at low $V_{sd}$ bias, though for very different $\Delta V_{sd}$ ranges. In particular, $\Delta V_{sd}$ decreases with decreasing $D$, i.e. $\Delta V^I_{sd}<\Delta V^{II}_{sd}<\Delta V^{III}_{sd}$. Correspondingly, the differential conductance ($dI/dV$) measured as a function of $V_{sd}$ for the same gate configurations is low when the Fermi level is in the middle of the band-gap. In contrast, a high $dI/dV$ is observed when the Fermi level is in the conduction or valence bands, see Fig. \ref{fig3}b.

To demonstrate that we probe an intrinsic band-gap in these electrical transport experiments, we performed a detailed study of the temperature dependence of the conductivity. These measurements can distinguish if charge carriers are hopping between sub-gap impurity states or if they are thermally excited between valence and conduction band over the energy gap ($\Delta$). In the hopping regime, depending on the specific conduction mechanism and on the relevance of correlation effects, the temperature dependence of conductivity ($\sigma$) for a two dimensional system is described by $\sigma = \sigma_0 exp(-(T_{0}/T)^{n})$ with $n= 1, 1/3$ or $1/2$ corresponding to nearest neighbour hopping, two-dimensional Mott variable range hopping and Efros-Shklovskii variable range hopping in the
presence of Coulomb interaction between the localised states. $T_{0}$ is a characteristic parameter which is a function of the activation energy of hopping \cite{Shloski}. On the other hand, in the absence of impurity states thermally activated electrical transport described by $\sigma=\sigma_0 exp(-\Delta/2 k_{B}T)$ ($k_{B}$ Boltzman constant) is expected. In addition, for short channel high quality devices and at low temperatures ($k_{B}T<\Delta$), a sub-gap electrical current is expected due to quantum tunneling through the energy barrier defined by $\Delta$.

\begin{figure}{}
\center
		\includegraphics[width=0.8\textwidth]{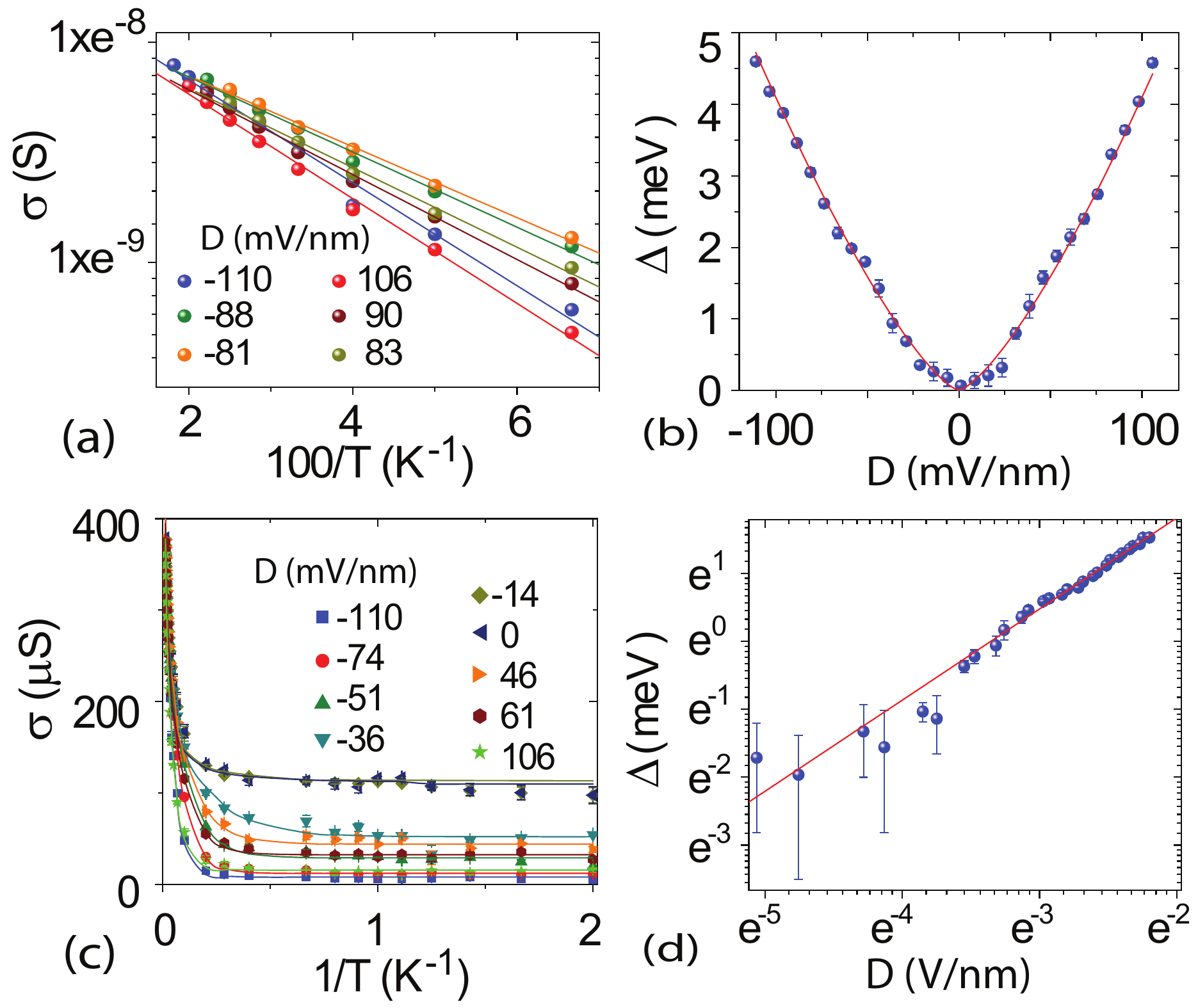}
\caption{\label{fig4} (a) Arrhenius plot for different values of external electric displacement. The dots are experimental data and the continuous lines are fit to thermally activated transport over the band-gap $\Delta$. $e$ on the conductivity axis is the constant of Napier (b) Shows the estimated values of the band-gap as a function of $D$. The continuous line is a fit to $\Delta \propto D^{3/2}$. (c) Plot of the minimum of conductivity as a function of inverse temperature for different average external electric displacements, as indicated. The dots are experimental data points, whereas the continuous line is a fit to Eq. 1. (d) Linear fit of $ln(\Delta)$ \textit{vs} $ln(D)$ with best fit $N=3 \pm 0.2$. $e$ on the axis of $\Delta$ is the constant of Napier.}
\end{figure}

Figure \ref{fig4}a shows some representative Arrhenius plots of the temperature dependence of the minimum conductivity $\sigma$ for different values of $D$. It is apparent that in all cases the experimental data are well described by a linear fit, as expected for thermally activated transport in intrinsic semiconductors and $\Delta$ is the only fitting parameter for the slope. We find that the band-gap can be continuously tuned up to $\Delta=4.6 \pm 0.06 meV$ at $D=-110 mV/nm$, see Fig. 4b. To understand the temperature dependence of $\sigma$ down to cryogenic temperatures (see Fig. 4c), we consider the combination of thermally activated transport and quantum tunneling through a short-channel semiconductor with a clean band-gap. The conductivity is given by the relation:

\begin{equation}
\sigma =2 \mu e \int_{\Delta/2}^{\infty} DOS(E) f(E) dE + \sigma_1,
\end{equation}

where $DOS(E)$ is the density of states for ABC-trilayer, $f(E)$ is the Fermi distribution and $\sigma_1$ is the measured low-temperature residual conductivity due to quantum tunneling (see supplementary information). The low-energy dispersion of ABC-trilayer is described by one pair of conduction and valence bands trigonally warped for $E<10meV$ evolving into a cubic dispersion ($E \propto k^3$) for $E>10meV$ \cite{Koshino2009,Koshino2010,Avetisyan2010,Zhang2010}. At $E=10meV$ the Fermi surface undergoes a topological discontinuity known as Lifshitz transition \cite{Koshino2009,Zhang2010,Avetisyan2010,Bao2011}. In our experiments, both the temperature range and the values of $\Delta$ (see Fig. 4b) are smaller than the energy corresponding to the Lifshitz transition ($E_{L}$). Therefore, we are probing electrical transport at the edges of the conduction and valence bands. The opening of a band-gap introduces a parabolicity at the band edges for each separate pocket known to exist in the low energy dispersion of ABC-trilayer \cite{Koshino2009}, resulting in a constant density of state equal to $m^*/2\pi \hbar^2$ with $m^*$ effective mass for each pocket. Figure \ref{fig4}c shows that considering a constant $DOS(E)$ in Eq. (1) gives an accurate fit to the experimental data using as the only fitting parameter $\Delta$ since the mobility is obtained from the field effect measurements, the effective mass ($m^*=0.05\pm 0.01 m_0$, with $m_0$ free electron mass) is kept fixed after the first fit to the data at D=100 mV/nm and $\sigma_1$ is the low temperature residual conductivity. For $\Delta \ll E_{L}$ the low energy dispersion is similar to 3 gapped Dirac cones (i.e. $DOS(E)\approx |E| \cdot \theta(|E|-\Delta|)$). However, even in this limit, the difference between the values of $\Delta$ found considering a linear or parabolic dispersion is within experimental error. The estimated values of $\Delta$ from the fit to Eq. (1) are consistent with the values found from the Arrhenius plot analysis and they also correspond to the energy scale of the non-linearity measured in the I-V characteristics (see Fig. \ref{fig3}a and Fig. \ref{fig3}b). The self-consistent tight binding model predicts that the low-electric field dependence of the energy gap in ABC-multilayers is $\Delta \propto D^{N/2}$, with $N$ the number of layers \cite{Koshino2010,Zhang2010}. We verify this theoretical prediction by fitting the experimental data for $ln(\Delta)$ $\textit{vs.}$ $ln(D)$ to a linear slope. The best fit gives $N=3 \pm 0.2$ as expected for the intrinsic band-gap in ABC-trilayer graphene, see Fig. \ref{fig4}d. 

Finally, we note that both Nearest Neighbour Hopping (NNH) and thermally activated transport have a similar exponential decay of the resistivity as a function of temperature \cite{Shloski,Withers2011}. However, in our devices the electron mean free path at the edge of the conduction and/or valence bands is typically $\geq 200nm$ -comparable to the transistor channel length and width (both $\sim 500nm$). Therefore, the electrical transport regime in our experiments is quasiballistic and not diffusive as in NNH. If we assume that $\Delta$ is the activation energy of NNH, we find a localization radius of 10nm, i.e. one order of magnitude smaller than the electron mean free path in these devices. Consistently, we find that the values of $\Delta$ estimated from $\sigma(T)$ correspond to the energy scale of the non-linearity measured in the I-V characteristics, see inset in Fig. \ref{fig3}a and Fig. \ref{fig3}b, contrary to NNH regime \cite{Taychatanapat2010,Withers2011}. Furthermore, we estimate the same values of $\Delta$ in devices with different aspect ratios (see supplementary information), conclusively demonstrating that our experiments provide a direct measurement of the gate-tunable band-gap in transport experiments in ABC-trilayer. 

In conclusion, we have reported a detailed study of the source-drain bias, temperature and perpendicular electric field dependence of the resistance in suspended double gated ABC-trilayer graphene. All these independent measurements consistently show the appearance of an energy gap in the density of states of ABC-trilayer. We demonstrate that we can tune $\Delta$ up to $4.6 meV$ for an average electrical displacement of $110 mV/nm$. Our experiments present the first direct observation for a clean gate-tunable band-gap in electrical transport measurements in ABC-trilayer graphene. These findings indicate that high quality few-layer graphene are an attractive system for the developement of gate tunable and highly efficient THz sources and detectors where a clean energy gap is an essential requirement.

\textbf{Acknowledgements.} S.R. and M.F.C acknowledge financial support from EPSRC (Grant no. EP/J000396/1, EP/K017160/1,
EP/K010050/1, EPG036101/1, EP/M001024/1, EPM002438/1), from Royal Society international
Exchanges Scheme 2012/R3 and 2013/R2 and from DSTL Quantum 2.0 Technologies. The authors knowledge useful discussions with E. Mariani.

\textbf{Supporting information.} Additional information on fabrication of suspended and doubly gated ABC-stacked trilayer graphene, current annealing of suspended and doubly gated ABC-stacked trilayer graphene, energy gap in other doubly gated ABC-stacked trilayer graphene devices and low temperature quantum tunnel conductivity is available in the Supporting Information. This material is available free of charge via the Internet at http://pubs.acs.org..

\end{document}